\documentclass[fleqn,10pt]{wlscirep}
\usepackage[utf8]{inputenc}
\usepackage[T1]{fontenc}
\title{Numerical Validation of a MOSFET-Based Control Circuit for High-Power Intelligent Reflecting Surfaces for Wireless Power Transfer Applications}

\author[1]{Gakuto Ichikawa}
\author[1]{Eisuke Omori}
\author[1]{Atsuko Nagata}
\author[2]{Akise Kumashiro}
\author[2]{Kazuhiro Kizaki}
\author[2]{Shunsuke Saruwatari}
\author[1,*]{Hiroki Wakatsuchi}
\affil[1]{Department of Engineering, Graduate School of Engineering, Nagoya Institute of Technology, Nagoya, Aichi 466-8555, Japan}
\affil[2]{Graduate School of Information Science and Technology, The University of Osaka, Suita, Osaka 565-0871, Japan}

\affil[*]{wakatsuchi.hiroki@nitech.ac.jp}

\begin{abstract}
Intelligent reflecting surfaces (IRSs) have attracted considerable attention because of their ability to dynamically control electromagnetic wave propagation. While most existing IRSs have been developed for low-power communication and sensing applications, their extension to high-power wireless power transfer (WPT) environments remains largely unexplored, as the high induced currents can damage or saturate the sensitive control elements, disrupting their tuning functionality. Here, we propose a metal-oxide-semiconductor field-effect transistor-based (MOSFET-based) binary control circuit for IRSs operating at 2.4 GHz that can withstand input power levels exceeding 1 W per unit cell. The control circuit employs a back-to-back MOSFET switching topology with series and parallel capacitors to suppress impedance variations arising from device nonlinearity while maintaining a reflection phase difference of approximately 180 degrees between the ON and OFF states. A theoretical model based on transmission lines is developed and validated against full-wave co-simulations incorporating nonlinear SPICE device models. The dynamic range is evaluated with respect to both the rated current and the reflection phase difference, demonstrating stable operation up to 1.25 W. Supercell-level beam steering is further demonstrated through far-field simulations, confirming active control of the reflection angle via switching pattern reconfiguration. These results establish a foundation for the deployment of IRSs in high-power WPT scenarios.
\end{abstract}
\begin{document}

\flushbottom
\maketitle
%
%
\thispagestyle{empty}

\section*{Introduction}
The rapid proliferation of Internet of Things (IoT) devices has fundamentally transformed modern society, enabling smart infrastructure across domains ranging from environmental monitoring and industrial automation to healthcare and smart agriculture\cite{gubbi2013internet, choudhary2024internet}. As the number of wirelessly connected devices continues to grow and each device requires a reliable source of electrical power, the conventional approach of wired power delivery and manual battery replacement becomes unrealistic. Wireless power transfer (WPT) has therefore emerged as a key enabling technology that can deliver energy through free space, eliminating physical cables and extending the operational lifetime of distributed sensor networks\cite{lu2015wireless, sasatani2021room, alabsi2024wireless}. Among the various WPT methods, near-field approaches based on electromagnetic induction are effective over short distances of several tens of centimetres, whereas far-field microwave-based WPT can transmit energy over distances of several metres or more, making it particularly attractive for powering spatially dispersed IoT nodes across large indoor or outdoor environments\cite{lu2015wireless, sasatani2021room, alabsi2024wireless, suzuki2024experimental, dang2026metasurface}.

A fundamental limitation of far-field WPT is that its transmission efficiency decreases sharply when obstacles---such as walls, furniture, or structural columns---obstruct the direct line-of-sight (LOS) path between the energy transmitter and the receiving devices. In such non-line-of-sight (NLOS) scenarios, the microwave beam is partially reflected, scattered, or absorbed by the intervening objects, reducing the power density arriving at the receiver. While natural reflections from walls and ceilings can redirect a portion of the transmitted energy, the reflection direction is governed solely by the geometry and material properties of the reflecting surface and cannot readily be controlled. Consequently, devices located in shadowed regions behind obstacles cannot be efficiently powered without an auxiliary mechanism that actively reshapes the propagation environment.

Metasurfaces---two-dimensional artificial electromagnetic materials composed of subwavelength periodic unit cells---offer an elegant solution to this problem\cite{EBGdevelopment, MTMbookEngheta, yu2014flat}. Unlike conventional materials whose electromagnetic properties are dictated by their molecular composition and are therefore difficult to engineer at will, metasurfaces can derive their effective properties from the geometry of metallic elements patterned on a dielectric substrate\cite{smith2005electromagnetic, holloway2005reflection, pfeiffer2013metamaterial, fathnan2022method}. When an electromagnetic wave impinges on such a surface, the electric charges induced on the metallic patches oscillate at a resonant frequency determined by the unit cell dimensions. By adjusting structural parameters and composite media, the effective permittivity and permeability can be designed to achieve a wide range of values, including negative refractive indices, i.e., properties not found in nature\cite{smithDNG1D, smithDNG2D2}. This design freedom has spawned a rich landscape of electromagnetic functionalities: perfect lenses that overcome the diffraction limit\cite{pendryperfetLenses, ZhangHyperLens}, ultrathin absorbers\cite{mtmAbsPRLpadilla, My1stAbsPaper}, electromagnetic cloaking devices that render objects invisible to incident radiation\cite{schurig2006metamaterial, pendryCloaking}, and phase-gradient surfaces that redirect reflected and transmitted wavefronts according to the generalised Snell's law\cite{yu2011light}.

Among these functionalities, electronically reconfigurable wavefront shaping is of particular relevance to WPT. An intelligent reflecting surface (IRS), also referred to as a reconfigurable intelligent surface (RIS), is a planar metasurface in which each unit cell contains a tuneable element whose impedance can be switched or varied by an external control signal\cite{yu2011light, pfeiffer2013metamaterial}. By programming the spatial distribution of reflection phases across the aperture, the IRS can steer an incident beam towards an arbitrary direction\cite{yu2011light, omori2026experimental}, split a single beam into multiple beams directed at different receivers\cite{cui2014coding, zhang2018space}, or focus energy onto a specific spot\cite{zhou2020intelligent, fathnan2022method}. Significant advances have been reported in the context of wireless communications, where IRSs enhance signal coverage and capacity in 5G and beyond-5G networks by creating virtual LOS links in environments with optimised scattering\cite{zhou2020intelligent}. Furthermore, IRSs have been investigated for use in radar sensing\cite{shao2022target} and near-field imaging\cite{jiang2024near}, where the ability to generate programmable illumination patterns enables high-resolution target reconstruction.

\begin{figure}[tb!]
\centering
\includegraphics[width=\linewidth]{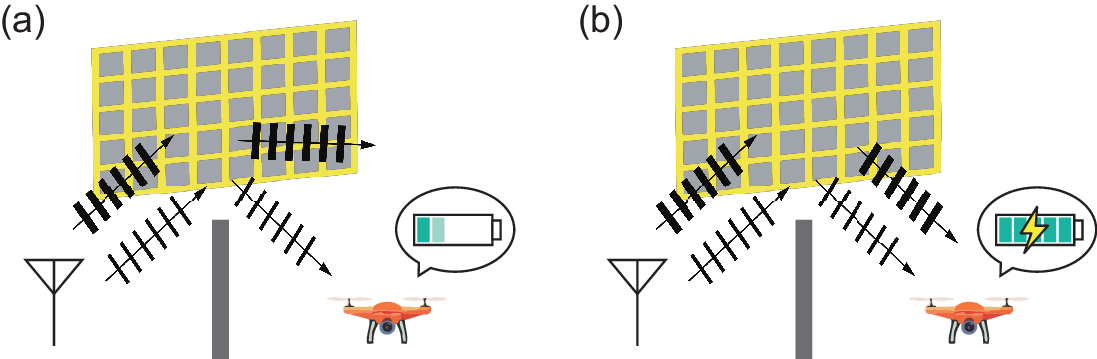}
\caption{The conceptual image of the proposed IRS. (a) A conventional IRS operating only for small-signal wireless communications. (b) The proposed IRS operating for both small-signal wireless communications and high-power WPT. }
\label{fig:1}
\end{figure}

Despite these advances, nearly all existing IRS implementations rely on nonlinear tuneable circuit components, such as PIN diodes and varactor diodes. PIN diodes offer fast switching between two impedance states\cite{li2017nonlinear} and are thus well suited to binary-coded metasurfaces\cite{cui2014coding, huang2022switchable}, whereas varactors provide continuous capacitance tuning, enabling multibit or analogue phase control. However, both device types are designed for small-signal microwave applications and possess limited current-handling capability---typically on the order of tens to hundreds of milliamperes (Fig. \ref{fig:1}a). When the incident power density increases to the levels encountered in microwave WPT, the incident wave directly interferes with these tuneable elements, which makes it difficult to adjust the sensitive circuit elements (and the impedances of the entire metasurface cells) via the equipped control circuitry. Additionally, the induced surface currents on the metasurface can easily exceed these ratings, leading to irreversible device failure. For this reason, the extension of IRS technology from the milliwatt-class communication regime to the watt-class WPT regime is an open and significant challenge.

To bridge this gap, we propose the use of metal--oxide--semiconductor field-effect transistors (MOSFETs) as the switching elements within an IRS designed for high-power operation at 2.4 GHz (Fig. \ref{fig:1}b). MOSFETs are voltage-controlled devices in which a channel between the source and drain terminals is formed or depleted by the gate--source voltage, enabling ON/OFF switching of the current flow. Compared with PIN diodes, commercially available MOSFETs offer substantially higher pulse-rated currents (hundreds of milliamperes to several amperes), superior thermal stability, and mature, low-cost packaging. A single MOSFET, however, cannot block alternating current (AC) in both directions because of the intrinsic body diode, which conducts in the reverse direction. We therefore adopt a back-to-back source-connected dual-MOSFET topology in which the two body diodes oppose each other, ensuring that at least one diode is reverse-biased regardless of the instantaneous current polarity, thereby achieving bidirectional current blocking in the OFF state.

A further complication arises at gigahertz frequencies: even in the OFF state, the parasitic drain--source capacitance of each MOSFET leads to a low-impedance leakage path for the high-frequency current such that the switching circuit cannot act as an ideal open circuit. Instead, the circuit behaves as a tuneable impedance element whose ON-state impedance is dominated by the channel resistance and whose OFF-state impedance is dominated by the parasitic capacitance. Moreover, both the resistance and the capacitance are nonlinear functions of the voltage across the device, meaning that the effective impedance---and hence the reflection phase of the metasurface---depends on the incident power level. To address this nonlinearity, we introduce an auxiliary stabilisation circuit structure composed of a series capacitance and a parallel capacitance around the switching circuit. The parallel capacitor mitigates the dominant susceptance so that the MOSFET's impedance fluctuations do not severely influence the total circuit admittance. Meanwhile, the series capacitor is used to maintain a sufficient susceptance contrast between the ON and OFF states, ensuring that the binary reflection phase difference remains close to 180$^\circ$.

In this paper, we present the complete design, theoretical analysis, and full-wave numerical validation of the proposed MOSFET-based IRS. The equivalent circuit of the unit cell loaded with the control circuit is modelled as a shunt-impedance-loaded transmission line, from which the reflection coefficient in both switching states is derived analytically. This model is compared against co-simulations that couple a finite-element electromagnetic solver with a nonlinear circuit simulator incorporating the manufacturer's SPICE model of the selected MOSFET. Three design variants, differing in the shunt inductance used for frequency tuning, are evaluated in terms of the reflection phase, reflection magnitude, and input current, revealing a trade-off between nonlinearity suppression and reflection efficiency. The dynamic range of each design is quantified using two complementary criteria: the maximum input power at which the device current remains below the pulse-rated value and the minimum input power at which the binary reflection phase difference satisfies the operational threshold. Finally, beam-steering functionality is demonstrated at the supercell level through far-field radiation-pattern analysis of impedance-sheet models derived from the co-simulation data. The results confirm that the proposed IRS architecture supports active beam control at input powers exceeding 1 W per unit cell at 2.4 GHz, substantially surpassing the operating regime of conventional diode-based IRSs.

\section*{Results}
\subsection*{Impedance characterisation of the MOSFET switching circuit}

\begin{figure}[htb!]
\centering
\includegraphics[width=\linewidth]{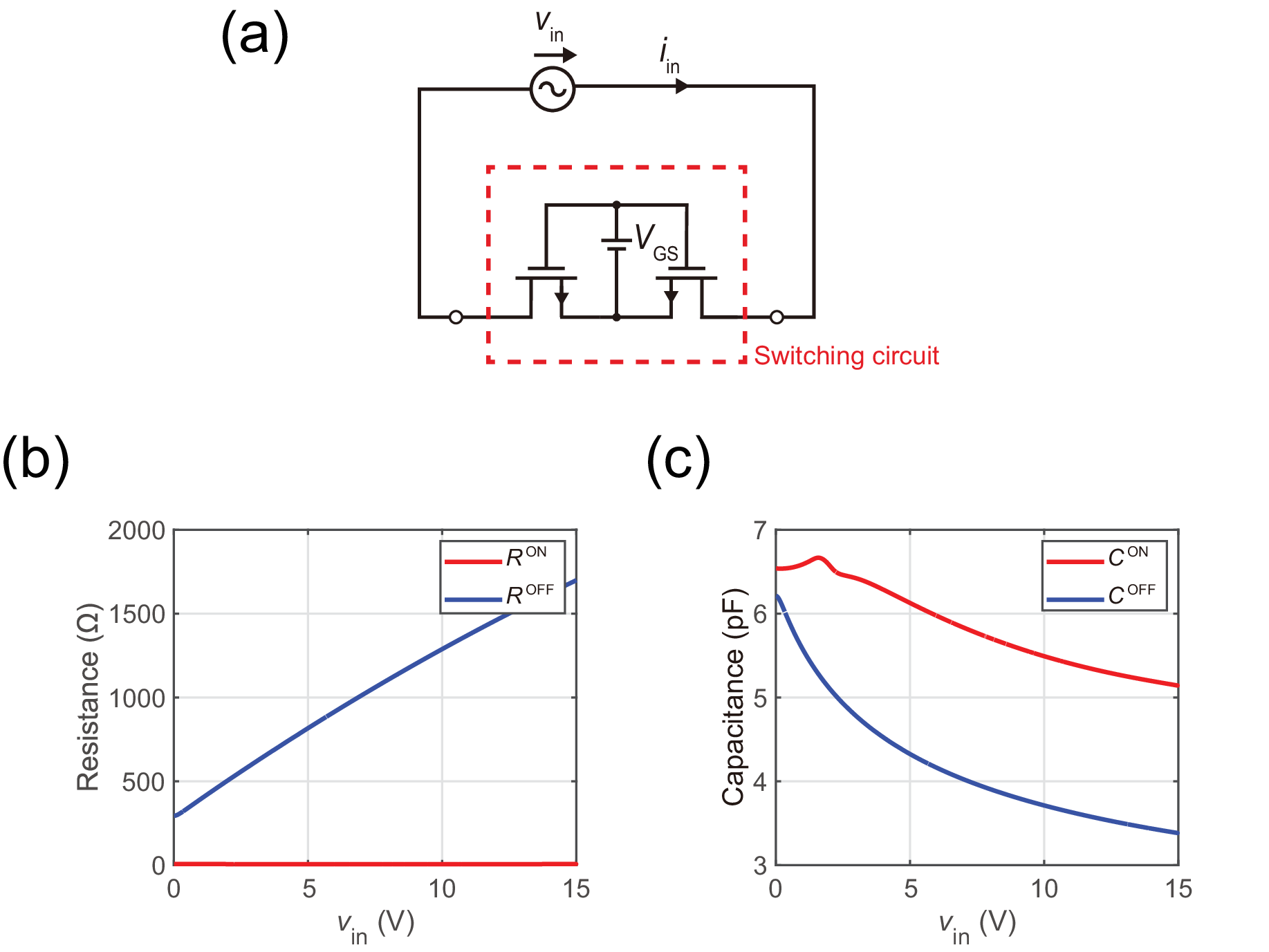}
\caption{Equivalent parameters of the MOSFET switching circuit extracted at 2.4 GHz as a function of the input voltage $v_\mathrm{in}$. (a) The MOSFET switching circuit. (b) The parallel resistance and (c) parallel capacitance for the ON and OFF states.}
\label{fig:2}
\end{figure}

The fundamental switching circuit design (Fig. \ref{fig:2}a) was constructed from two Toshiba SSM3K72CFS n-channel MOSFETs selected for their low parasitic output capacitance ($C_{oss}$ < 7 pF), which is critical for maximising the impedance contrast at 2.4 GHz. The two MOSFETs were connected in a back-to-back configuration with their source terminals tied together so that the body diodes face in opposing directions. This topology ensures that regardless of the instantaneous polarity of the alternating induced current, one of the two body diodes is always reverse-biased, thereby preventing unwanted through-conduction in the OFF state.

To extract the equivalent impedance of the switching circuit, a sinusoidal voltage source at 2.4 GHz was applied across the drain terminals, and the resulting current was computed via a harmonic balance analysis using the manufacturer's PSPICE model. The gate--source voltage $V_{GS}$ was set to 5 V for the ON state and 0 V for the OFF state. From the input voltage and current phasors, the series-equivalent resistance $R_s$ and reactance $X_s$ were obtained and subsequently converted to a parallel $RC$ representation using the standard series-to-parallel transformation. The resistive component is given by $R = (R_s^2 + X_s^2)/R_s$, and the capacitive component is given by $C = -X_s/(\omega(R_s^2 + X_s^2))$, where $\omega = 2\pi f$ is the angular frequency and $f$ represents the frequency.

The extracted equivalent resistance and capacitance are plotted as functions of the input voltage amplitude in Fig. \ref{fig:2}b and Fig. \ref{fig:2}c. Note that in this study, the parameters or variables associated with the OFF and ON states are represented by the superscripts OFF and ON, respectively (e.g., as shown by $R^{\mathrm{OFF}}$ and $R^{\mathrm{ON}}$ in Fig. \ref{fig:2}b). In the OFF state, the channel is pinched off, and the resistance $R^{\mathrm{OFF}}$ exceeds 1 k$\Omega$ at low voltage levels, reflecting the high impedance of the depleted channel. However, at 2.4 GHz, the reactance of the parasitic capacitance $C^{\mathrm{OFF}}\approx$ 3.6 pF is only approximately 18 $\Omega$, which is far lower than $R^{\mathrm{OFF}}$ and therefore dominates the total impedance. This means that even in the nominally  off  state, a significant fraction of the incident current leaks through the parasitic capacitance---a behaviour fundamentally different from that of an ideal switch and one that must be explicitly accounted for in the metasurface design. In the ON state, the formation of the inversion layer reduces the channel resistance to $R^{\mathrm{ON}}\approx$ 5 $\Omega$, whereas the capacitance increases to $C^{\mathrm{ON}}\approx$ 6.4 pF. Current therefore flows through both the resistive channel and the capacitive path in parallel. 

\begin{figure}[ht!]
\centering
\includegraphics[width=0.5\linewidth]{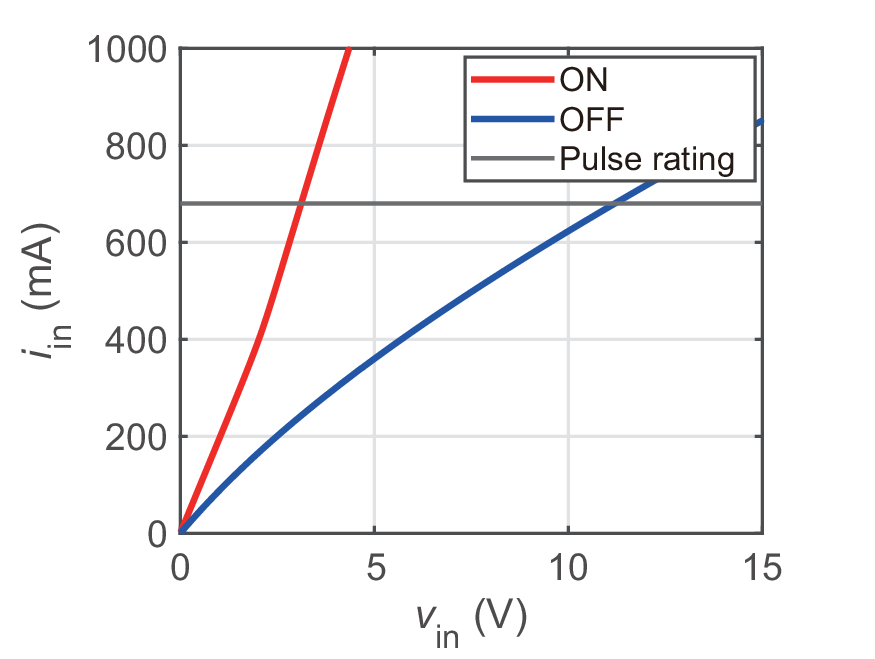}
\caption{Input current $i_\mathrm{in}$ of the switching circuit as a function of the input voltage $v_\mathrm{in}$ for the ON and OFF states, compared with the pulse-rated current of 680 mA.}
\label{fig:3}
\end{figure}

Fig. \ref{fig:3} compares the peak current flowing through the switching circuit with the pulse-rated current of 680 mA specified in the device datasheet. Because the ON-state impedance is much lower than the OFF-state impedance, the ON-state current exceeds the OFF-state current at any given input voltage, and it is the ON-state current that limits the maximum tolerable input voltage. Specifically, the ON-state current reaches 680 mA at an input voltage of $v_{\mathrm{in}}$ = 3.11 V, whereas the OFF-state current does not reach this threshold until $v_{\mathrm{in}}$ = 11.20 V. To ensure safe operation, all subsequent designs use the equivalent parameters evaluated at the ON- and OFF-state rated-current boundaries, as summarised in Table \ref{tab:1}. These values represent the worst-case operating point in terms of device stress and therefore yield conservative design margins.

\begin{table}
    \centering
    \begin{tabular}{ccccc}
    \hline
      Parameter   & Min & Max & Average & At rated limit\\
      \hline \hline 
      $R^{\mathrm{ON}}$ ($\Omega$)  & 5.02 & 6.05 & 5.74 & 5.02\\
      $R^{\mathrm{OFF}}$ ($\Omega$)  & 294 & 1392 & 864 & 1390\\
      $C^{\mathrm{ON}}$ (pF)  & 6.41 & 6.67 & 6.55 & 6.41\\
      $C^{\mathrm{OFF}}$ (pF)  & 3.62 & 6.21 & 4.42 & 3.61\\\hline
    \end{tabular}
    \caption{Equivalent parameters of the MOSFET switching circuit at the rated-current boundary.}
    \label{tab:1}
\end{table}

\subsection*{Control circuit design for nonlinearity suppression and phase contrast}
\begin{figure}[ht!]
\centering
\includegraphics[width=\linewidth]{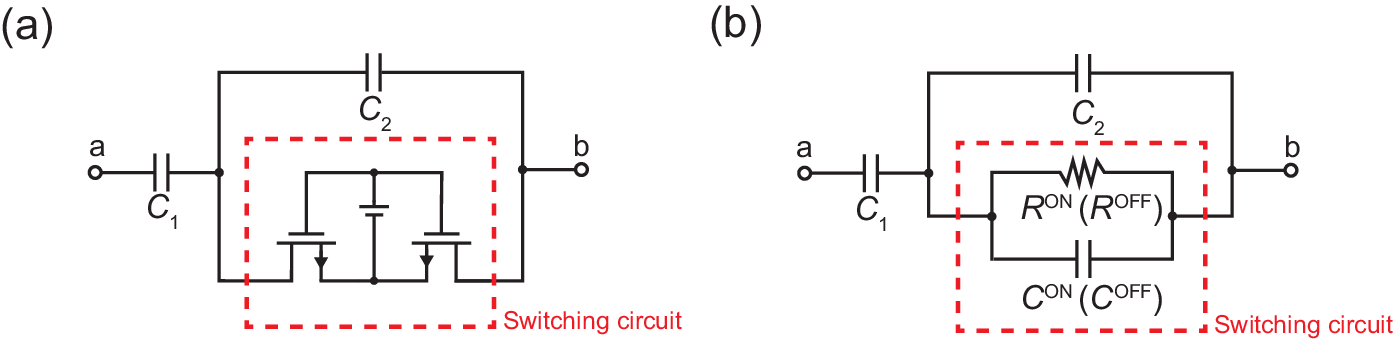}
\caption{Proposed control circuit. (a) Circuit topology showing the series capacitance $C_1$, parallel capacitance $C_2$, and back-to-back MOSFET switching circuit. (b) Equivalent circuit model with the MOSFET represented as a parallel RC network.}
\label{fig:4}
\end{figure}

\begin{figure}[ht!]
\centering
\includegraphics[width=\linewidth]{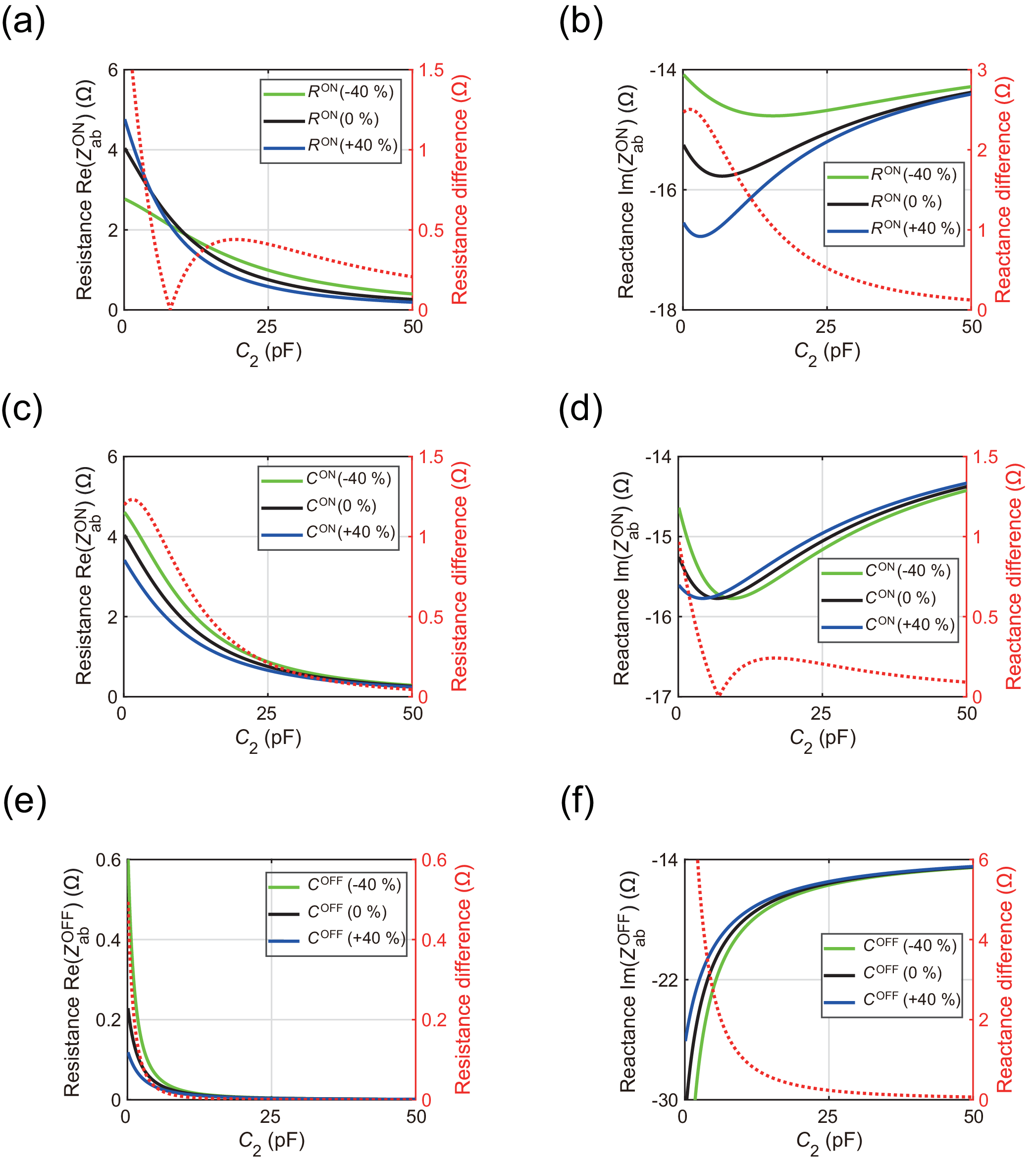}
\caption{Suppression of the impedance variation by the parallel capacitance $C_2$. (a-f) Resistance and reactance of the control circuit as a function of $C_2$ when (a,b) $R^{\mathrm{ON}}$, (c,d) $C^{\mathrm{ON}}$, and (e,f) $C^{\mathrm{OFF}}$ are perturbed by $\pm$40\%. Solid curves (left axes) show $\operatorname{Re}(Z_\mathrm{ab})$ or $\operatorname{Im}(Z_\mathrm{ab})$ for the $-$40\%, nominal, and $+$40\% cases, whereas the red dotted curves (right axes) show the spread between the $-$40\% and $+$40\% cases.}
\label{fig:5}
\end{figure}

As demonstrated above, the MOSFET switching circuit exhibits pronounced voltage-dependent impedance variations arising from the nonlinear relationship between the drain--source voltage and the channel conductance and capacitance. If the bare switching circuit were inserted directly into the unit cell gap, the reflection phase would vary appreciably with the incident power, undermining the predictability and stability of the beam-steering function of the IRSs. To mitigate this issue and simultaneously ensure a sufficiently large binary phase contrast, we introduce an auxiliary stabilisation network consisting of a series capacitance $C_1$ and a parallel capacitance $C_2$ connected around the switching circuit, as shown in Fig. \ref{fig:4}. The impedances of the resulting control circuit in the ON and OFF states, $Z_{ab}^{\mathrm{ON}}$ and $Z_\mathrm{ab}^{\mathrm{OFF}}$, respectively, are given by
\begin{eqnarray}
Z_\mathrm{ab}^{\mathrm{ON}} &=& \cfrac{1}{j\omega C_1} +\cfrac{1}{\cfrac{1}{R^{\mathrm{ON}}} + j\omega(C^{\mathrm{ON}} + C_2)},\\
Z_\mathrm{ab}^{\mathrm{OFF}} &=& \cfrac{1}{j\omega C_1}+ \cfrac{1}{\cfrac{1}{R^{\mathrm{OFF}}} + j\omega(C^{\mathrm{OFF}} + C_2)}.   
\end{eqnarray}
The role of $C_2$ is to provide a large baseline susceptance that mitigates the impedance fluctuations of the switching circuit. To quantify this stabilisation effect, we fix $C_1$ = 5 pF and independently sweep the three voltage-sensitive MOSFET parameters---$R^{\mathrm{ON}}$, $C^{\mathrm{ON}}$, and $C^{\mathrm{OFF}}$---by $\pm$40\% from their nominal values, simulating worst-case manufacturing tolerances and power-level deviations. Figs. \ref{fig:5}a--f show the resulting resistance and reactance of the control circuit as functions of $C_2$. In any case, increasing $C_2$ eventually narrows the spread between the $-$40\% and $+$40\% curves, confirming that the parallel capacitor effectively lowers the influence of internal device variations. Physically, this occurs because a large $C_2$ lowers the impedance of the parallel branch so that the switching circuit's impedance fluctuation constitutes a progressively smaller fraction of the total admittance.

\begin{figure}[ht!]
\centering
\includegraphics[width=0.5\linewidth]{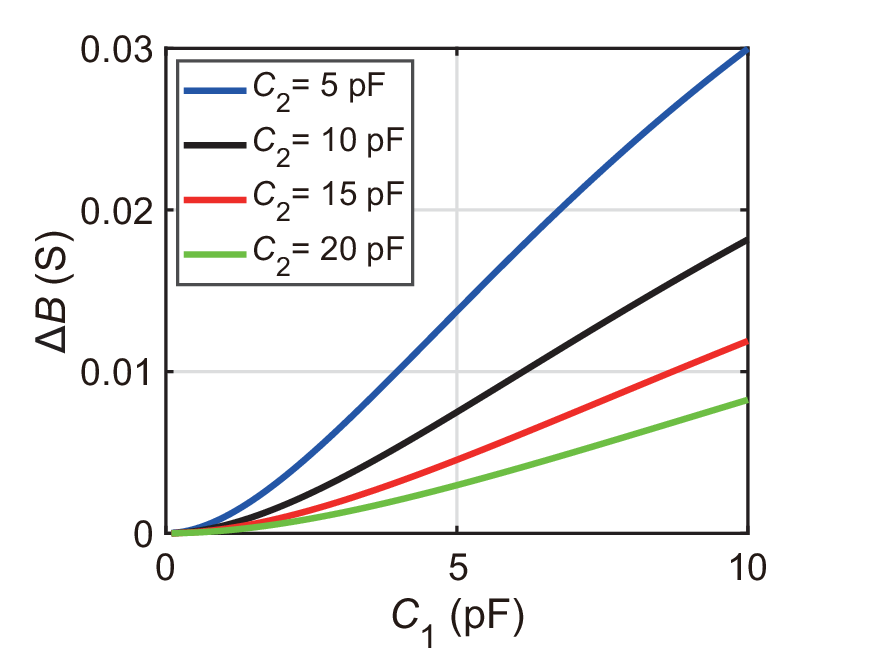}
\caption{Susceptance difference $\Delta B$ between the ON and OFF states as a function of the series capacitance $C_1$ for several values of $C_2$.}
\label{fig:6}
\end{figure}

However, the improved robustness afforded by a large $C_2$ comes at the cost of a reduced susceptance contrast $\Delta B$ between the ON and OFF states. The susceptance difference is defined as $\Delta B = |\operatorname{Im}(Y_\mathrm{ab}^{\mathrm{ON}}) - \operatorname{Im}(Y_\mathrm{ab}^{\mathrm{OFF}})|$, where $Y_\mathrm{ab} = 1/Z_\mathrm{ab}$ is the control circuit admittance and $\operatorname{Im}$ indicates the extraction of the imaginary component. As shown in Fig. \ref{fig:6}, for a fixed $C_1$, increasing $C_2$ progressively decreases $\Delta B$ because the change in the impedance of the switching circuit becomes an increasingly smaller perturbation of the total admittance. Since a binary-coded metasurface requires a reflection phase difference of approximately 180$^\circ$ between its two states, a minimum $\Delta B$ must be maintained. To enhance the reduced susceptance contrast, the series capacitance $C_1$ must be increased with $C_2$. In summary, the design rule for the proposed control circuit is to set both $C_1$ and $C_2$ as large as practically feasible: $C_2$ for nonlinearity suppression and $C_1$ for phase-contrast recovery.

\subsection*{Unit cell design and resonance tuning}
\begin{figure}[ht!]
\centering
\includegraphics[width=\linewidth]{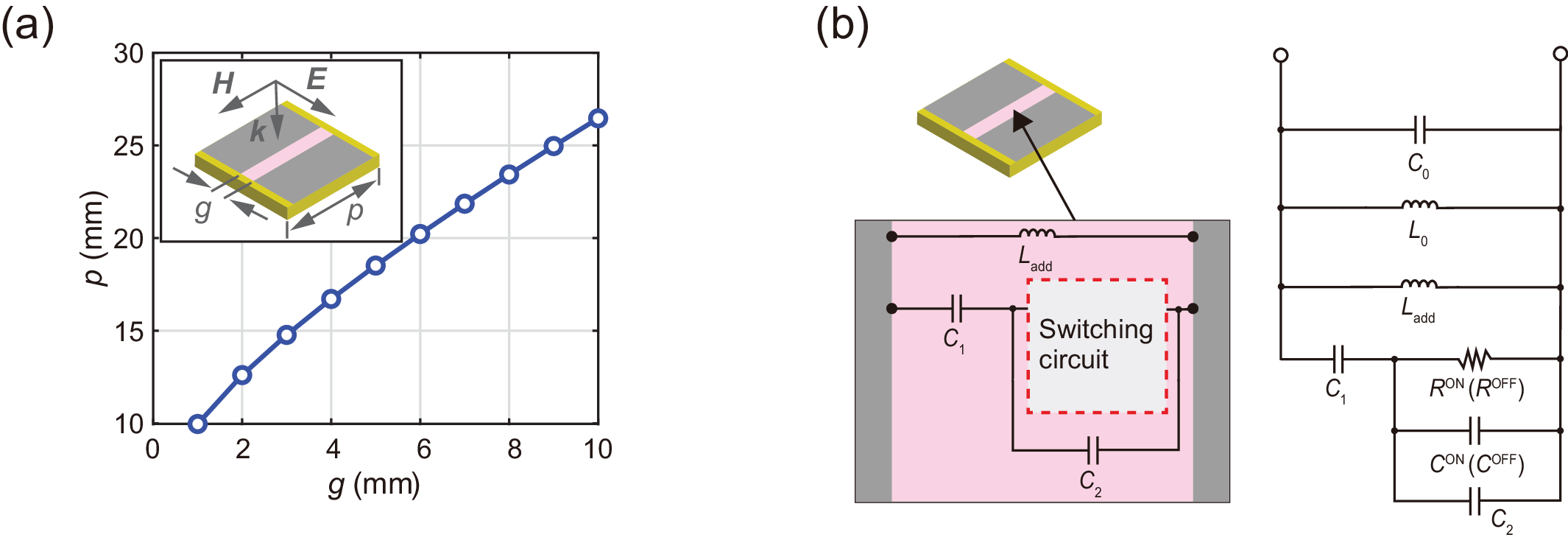}
\caption{Unit cell with an embedded control circuit. (a) The periodic unit cell without the loaded circuit (inset) and its period $p$ and gap $g$ satisfying the resonance condition. (b) Equivalent circuit of the loaded unit cell, including $C_0$, $L_0$, $L_{\mathrm{add}}$, and the control circuit.}
\label{fig:7}
\end{figure}

The unit cell follows a typical reflective patch metasurface architecture: a square metallic patch of width $w$ is designed on one face of a dielectric substrate (Rogers RO3003 with relative permittivity $\epsilon_{\mathrm{r}}$ = 3.00 and loss tangent tan $\delta$ = 0.0013), while a fully metallic ground (GND) plane covers the opposite face, as drawn in the inset of Fig. \ref{fig:7}a. The other structural parameters include the unit cell period $p$, the gap $g$ between patches, and the substrate thickness $h$. In the equivalent-circuit representation, the charge accumulation across the gap produces a lumped capacitance $C_0 = ((p - g)\epsilon_0(1 + \epsilon_{\mathrm{r}})/\pi) \cosh^{-1}{(p/g)}$ with $\epsilon _0$ representing the permittivity of vacuum, and the round-trip propagation delay through the grounded substrate gives an inductance $L_0 = \mu_0h$ with $\mu_0$ representing the permeability of vacuum \cite{sievenpiperPhD, luukkonen2009thin}. The resonant frequency $f_0$ of the unloaded unit cell is $f_0 = 1/(2\pi\sqrt{L_0C_0})$. Additionally, the proposed switching circuit is loaded across the patch gap with a composite capacitance $C_{\mathrm{add}}$. Assuming that $C_{\mathrm{add}}\approx C_1$ (i.e., the series capacitor is dominant), the resonant frequency shifts to $f_0 = 1/(2\pi\sqrt{L_0(C_0 + C_{\mathrm{add}})})$. Maintaining the target operating frequency of 2.4 GHz with a large $C_{\mathrm{add}}$ therefore requires the intrinsic cell parameters to be adjusted to smaller values, which constrains the geometric design space. Through parametric calculations, we found that $C_{\mathrm{add}}$ = 2 pF is the maximum value for which a physically realisable unit cell can be designed at 2.4 GHz. At this capacitance, the initial parameter set satisfying the resonance condition for a substrate thickness of $h$ = 1.52 mm is $p$ = 15 mm and $g$ = 3 mm (Fig. \ref{fig:7}a). However, electromagnetic simulations revealed significant capacitive coupling between the top patch and the ground plane at $h$ = 1.52 mm, which shifted the resonance away from the predicted value. To alleviate this coupling, the substrate thickness was increased to $h$ = 3.04 mm. Because no physically realisable pair ($p$, $g$) could satisfy the resonance condition at the greater thickness, the cell dimensions were retained at $p$ = 15 mm and $g$ = 3 mm, and an additional shunt inductance $L_{\mathrm{add}}$ was introduced in parallel with the control circuit to restore the operating frequency (Fig. \ref{fig:7}b). Thus, $f_0$ was modified to
\begin{eqnarray}
f_0 = \frac{1}{2\pi\sqrt{\cfrac{L_0L_{\mathrm{add}}}{L_0+L_{\mathrm{add}}}(C_0 + C_{\mathrm{add}})}}.
\label{eq:f0}
\end{eqnarray}
These final structural parameters are listed in Table \ref{tab:2}.

\begin{table}
    \centering
    \begin{tabular}{cc}
    \hline
       Parameter  & Value (mm) \\\hline \hline
       Unit cell period $p$ & 15.0 \\
       Gap width $g$ & 3.0 \\
       Substrate thickness $h$ & 3.04 \\
       Patch width $w$ & 12.0 \\\hline
    \end{tabular}
    \caption{Unit cell structural parameters.}
    \label{tab:2}
\end{table}

\subsection*{Theoretical design of binary reflection phases}
The complete loaded unit cell is represented by a shunt-impedance-loaded transmission line in which the free-space characteristic impedance $Z_0 = 120\pi$ $\Omega$ feeds the parallel combination of the unit cell admittance ($C_0$ and $L_0$) and the control circuit admittance ($C_1$, $C_2$, $L_{\mathrm{add}}$, and the MOSFET RC network), as depicted in the circuit model of Fig. \ref{fig:7}b. The input impedances in the ON and OFF states, represented by $Z_{\mathrm{in}}^{\mathrm{ON}}$ and $Z_{\mathrm{in}}^{\mathrm{OFF}}$, respectively, are obtained from the parallel sum of all admittance branches: 
\begin{eqnarray}
Z_{\mathrm{in}}^{\mathrm{ON}}
&=& \left[
  j\omega C_0
  + \frac{1}{j\omega L_0}
  + \frac{1}{j\omega L_{\mathrm{add}}}
  + \left(
      \frac{1}{j\omega C_1}
      + \frac{1}{\dfrac{1}{R_{\mathrm{ON}}} + j\omega(C_{\mathrm{ON}}+C_2)}
    \right)^{-1}
\right]^{-1},
\label{eq:ZinON}\\
Z_{\mathrm{in}}^{\mathrm{OFF}}
&=& \left[
  j\omega C_0
  + \frac{1}{j\omega L_0}
  + \frac{1}{j\omega L_{\mathrm{add}}}
  + \left(
      \frac{1}{j\omega C_1}
      + \frac{1}{\dfrac{1}{R_{\mathrm{OFF}}} + j\omega(C_{\mathrm{OFF}}+C_2)}
    \right)^{-1}
\right]^{-1}.
\label{eq:ZinOFF}
\end{eqnarray}
By using these input impedances, the corresponding reflection coefficients $\varGamma$s are obtained by
\begin{eqnarray}
\varGamma^{\mathrm{ON}}(C_1,C_2)
&=& \frac{Z_{\mathrm{in}}^{\mathrm{ON}} - Z_0}{Z_{\mathrm{in}}^{\mathrm{ON}} + Z_0}\nonumber\\
&=& \frac{\left[
  j\omega C_0
  + \cfrac{1}{j\omega L_0}
  + \cfrac{1}{j\omega L_{\mathrm{add}}}
  + \left(
      \cfrac{1}{j\omega C_1}
      + \cfrac{1}{\cfrac{1}{R_{\mathrm{ON}}} + j\omega(C_{\mathrm{ON}}+C_2)}
    \right)^{-1}
\right]^{-1} - Z_0}{\left[
  j\omega C_0
  + \cfrac{1}{j\omega L_0}
  + \cfrac{1}{j\omega L_{\mathrm{add}}}
  + \left(
      \cfrac{1}{j\omega C_1}
      + \cfrac{1}{\cfrac{1}{R_{\mathrm{ON}}} + j\omega(C_{\mathrm{ON}}+C_2)}
    \right)^{-1}
\right]^{-1} + Z_0},\label{eq:GammaON}\\
\varGamma^{\mathrm{OFF}}(C_1,C_2)
&=& \frac{Z_{\mathrm{in}}^{\mathrm{OFF}} - Z_0}{Z_{\mathrm{in}}^{\mathrm{OFF}} + Z_0}\nonumber\\
&=&\frac{\left[
  j\omega C_0
  + \cfrac{1}{j\omega L_0}
  + \cfrac{1}{j\omega L_{\mathrm{add}}}
  + \left(
      \cfrac{1}{j\omega C_1}
      + \cfrac{1}{\dfrac{1}{R_{\mathrm{OFF}}} + j\omega(C_{\mathrm{OFF}}+C_2)}
    \right)^{-1}
\right]^{-1} - Z_0}{\left[
  j\omega C_0
  + \cfrac{1}{j\omega L_0}
  + \cfrac{1}{j\omega L_{\mathrm{add}}}
  + \left(
      \cfrac{1}{j\omega C_1}
      + \cfrac{1}{\dfrac{1}{R_{\mathrm{OFF}}} + j\omega(C_{\mathrm{OFF}}+C_2)}
    \right)^{-1}
\right]^{-1} + Z_0}.
\label{eq:GammaOFF}
\end{eqnarray}
Here we first considered three candidate reflection phase pairs, each yielding a nominal 180$^\circ$ phase difference, more specifically, Design A with $\angle \varGamma^{\mathrm{ON}} = -90^\circ$ and $\angle \varGamma^{\mathrm{OFF}} = +90^\circ$, Design B with $\angle \varGamma^{\mathrm{ON}} = -135^\circ$ and $\angle \varGamma^{\mathrm{OFF}} = +45^\circ$, and Design C with $\angle \varGamma^{\mathrm{ON}} = -180^\circ$ and $\angle \varGamma^{\mathrm{OFF}} = 0^\circ$. Under these circumstances, we fixed the inductance $L_{\mathrm{add}}$ at 3.5 nH. In contrast, $C_1$ and $C_2$ values were derived using Eqs. (\ref{eq:GammaON}) and (\ref{eq:GammaOFF}). However, note that these equations cannot be arranged to directly calculate $C_1$ and $C_2$. Therefore, we swept both $C_1$ and $C_2$ from $0$~pF to $50$~pF in $0.01$~pF steps and minimised the deviation from the target reflection phases. In this search, computing the angular difference could lead to overestimation of the error near the $\pm180^\circ$ boundary. To address this issue, the error functions for the ON and OFF states were defined as the Euclidean distance $r$ on the unit circle in the complex plane:
\begin{eqnarray}
r^{\mathrm{ON}}
&=& \left| e^{j\angle\varGamma^{\mathrm{ON}}(C_1,C_2)}
       - e^{j\angle\varGamma_{\mathrm{target}}^{\mathrm{ON}}} \right|,
\label{eq:EON}\\
r^{\mathrm{OFF}}
&=& \left| e^{j\angle\varGamma^{\mathrm{OFF}}(C_1,C_2)}
       - e^{j\angle\varGamma_{\mathrm{target}}^{\mathrm{OFF}}} \right|,
\label{eq:EOFF}
\end{eqnarray}
where $\varGamma_{\mathrm{target}}$ represents the target reflection coefficient. We then found the $C_1$-$C_2$ pair that minimises a combined cost function $r_{\mathrm{total}}$ defined by 
\begin{eqnarray}
r_{\mathrm{total}} = r_{\mathrm{ON}} + r_{\mathrm{OFF}}.\label{eq:cost}
\end{eqnarray}
Finally, the optimal $C_1$ and $C_2$ were substituted back into Eqs. \eqref{eq:GammaON} and \eqref{eq:GammaOFF} to compute the theoretical reflection magnitude and phase of the unit cell. 

These theoretical results are shown in Table \ref{tab:3}. Designs A and B successfully achieved the target 180$^\circ$ phase difference. Design C, however, showed a reduced phase difference: the ON-state phase does not reach $-$180$^\circ$ because this limit corresponded to an infinitely large loading capacitance, which the finite susceptance contrast $\Delta B$ of the control circuit could not realize. Physically, a reflection phase of $-$180$^\circ$ would require the metasurface input impedance to approach zero (short-circuit condition), which necessitates an infinitely large shunt capacitance to cancel the inductive admittance of the substrate. Regarding the reflection magnitude, Design A exhibited almost perfect absorption in the ON state ($|\varGamma ^{\mathrm{ON}}|^2 \approx 0$) because the loaded unit cell impedance matched the free-space impedance, dissipating all incident energy within the metasurface design. Design C, while avoiding this resonance-matching problem, suffered from the aforementioned phase shortfall. Design B showed performance in between: it preserved the full 180$^\circ$ phase difference while maintaining a substantially higher ON-state reflection magnitude than Design A. We therefore selected Design B as the baseline configuration for all subsequent evaluations.

\begin{table}
    \centering
    \begin{tabular}{cccccccccc}
    \hline
     Design    & $\angle \varGamma^{\mathrm{ON}}_{\mathrm{target}}$ ($^\circ$) & $\angle \varGamma^{\mathrm{OFF}}_{\mathrm{target}}$ ($^\circ$) & $C_1$ (pF) & $C_2$ (pF)& $\angle \varGamma^{\mathrm{ON}}$ ($^\circ$) & $\angle \varGamma^{\mathrm{OFF}}$  ($^\circ$)& $|\varGamma^{\mathrm{ON}}|^2$ & $|\varGamma^{\mathrm{OFF}}|^2$\\\hline\hline
     A    & $-$90 & $+$90 & 2.34 & 6.96&$-$90.02&$+$91.73&0.0122&0.9829\\
     B    & $-$135 & $+$45 & 2.61 & 5.41&$-$134.92&$+$44.93&0.2940&0.9545\\
     C    & $-$180 & 0 & 4.95 & 0.02&$-$172.01&$-$0.10&0.8582&0.6968\\\hline
    \end{tabular}
    \caption{Target reflection phases and circuit constants for Designs A--C at 2.4 GHz ($L_{\mathrm{add}}$ = 3.5 nH).}
    \label{tab:3}
\end{table}

To further improve the reflection characteristics of Design B (with target phases fixed at $-135^\circ/+45^\circ$), we varied $L_{\mathrm{add}}$ as shown in Table \ref{tab:4} by using Eqs. \eqref{eq:GammaON} and \eqref{eq:GammaOFF}. Three values---$L_{\mathrm{add}} =$ 1.5, 2.5, and 3.5 nH---were assigned to Designs 1, 2, and 3, respectively. As $L_{\mathrm{add}}$ decreased, the required $C_2$ increased (Design 1: $C_2 =$ 19.36 pF versus Design 3: $C_2 =$ 5.41 pF), providing stronger nonlinearity suppression. At the same time, however, the larger $C_2$ reduced the ON-state reflection magnitude because the dominant parallel capacitance homogenised the impedance contrast and brought the ON-state impedance closer to free-space matching. This design space exploration clearly reveals the trade-off: Design 1 offers the greatest robustness against device variations but the lowest ON-state reflectivity, whereas Design 3 maximises reflectivity at the expense of reduced tolerance to nonlinearity. Design 2 provides an intermediate compromise. The implications of this trade-off for the overall dynamic range are quantified in the subsequent full-wave analysis.

\begin{table}[t]
\centering
\begin{tabular}{cccccccc}
\hline
Design & $L_{\mathrm{add}}$~(nH) & $C_1$~(pF) & $C_2$~(pF)  & $\angle\varGamma^{\mathrm{ON}}$ ($^\circ$) & $\angle\varGamma^{\mathrm{OFF}}$ ($^\circ$) & $|\varGamma^{\mathrm{ON}}|^2$ & $|\varGamma^{\mathrm{OFF}}|^2$\\
\hline\hline
1 & 1.5 & 4.41 & 19.36  & $-135.39$ & $+44.98$ & 0.0724 & 0.9736\\
2 & 2.5 & 3.15 &  9.16  & $-135.36$ & $+45.06$ & 0.1999 & 0.9645\\
3 & 3.5 & 2.61 &  5.41  & $-134.92$ & $+44.93$ & 0.2940 & 0.9545\\
\hline
\end{tabular}
\caption{Theoretical reflection magnitude and phase for Designs~1, 2, and~3 with different $L_{\mathrm{add}}$ values (target reflection phase: $-$135$^\circ$/$+$45$^\circ$).}
\label{tab:4}
\end{table}

\subsection*{Full-wave co-simulation and comparison with theoretical predictions}
To validate the analytical predictions and capture effects not included in the simplified equivalent-circuit model, a full-wave co-simulation was carried out using ANSYS Electronics Desktop 2025 R2. The unit cell was modelled in the finite-element solver HFSS with two pairs of periodic boundary conditions imposed on the lateral faces and a Floquet port placed at the top face, thereby emulating an infinite periodic array illuminated by a normal incident plane wave. A lumped port was assigned across the metallic gap to connect the loaded circuit elements in the circuit solver. The S parameters extracted from the HFSS model were imported into the ANSYS Circuit simulator, where the Toshiba SSM3K72CFS PSPICE model and the lumped elements ($C_1$, $C_2$, and $L_{\mathrm{add}}$) were connected at the lumped port. The gate--source voltage was set to 5 V (as the ON state) or 0 V (as the OFF state), and a harmonic balance analysis was performed at 2.4 GHz with the input power $P_{\mathrm{in}}$ swept from 0 to 1.5 W. The reflection coefficient was computed from the input and reflected voltage phasors at the Floquet port reference plane, with a phase correction of $\Delta \theta = 2k_0L$ applied to compensate for the round-trip propagation between the port and the metasurface. Here, $k_0$ denotes the wave vector in vacuum, while $L$ is the distance between the Floquet port and the metasurface ($L =$ 125 mm in this study). The structural and circuit parameters are summarised in Tables \ref{tab:2} and \ref{tab:4}.

\begin{figure}[ht!]
\centering
\includegraphics[width=\linewidth]{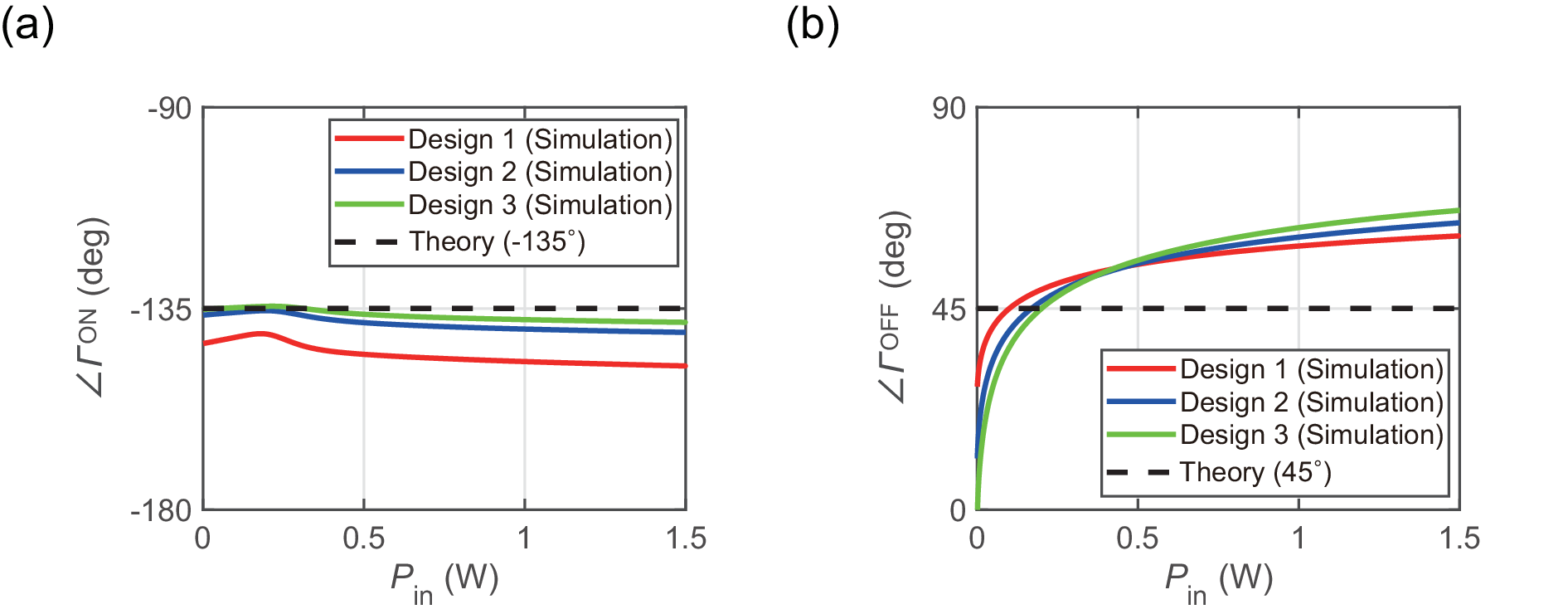}
\caption{Co-simulated reflection phase $\angle\varGamma$ versus input power $P_\mathrm{in}$ for Designs 1--3 in the (a) ON and (b) OFF states, compared with the theoretical target values.}
\label{fig:8}
\end{figure}

Fig. \ref{fig:8} presents the co-simulated reflection phase as a function of input power for Designs 1--3 in both the ON and OFF states. Unlike the analytical model, which uses linear circuit elements and therefore predicts power-independent reflection phases, the co-simulation reveals a clear power dependence attributable to the MOSFET's voltage-dependent impedance captured by the SPICE model. In the OFF state, the phase variation with power diminishes progressively from Design 3 to Design 1, confirming the stabilising role of the parallel capacitance $C_2$: Design 1 ($C_2 =$ 19.36 pF) suppresses the OFF-state phase variation most effectively because the large $C_2$ renders the MOSFET's impedance change insignificant relative to the total admittance. In the ON state, however, Design 1 exhibits the greatest phase variation despite having the largest $C_2$. This counterintuitive behaviour arises because the small $L_{\mathrm{add}} =$ 1.5 nH used in Design 1 amplifies the sensitivity of the overall input impedance to residual admittance perturbations: even though $C_2$ mitigates the absolute magnitude of the MOSFET's impedance change, the resulting small perturbation is magnified by the steep phase slope near the operating point set by the low-inductance branch.

\begin{figure}[ht!]
\centering
\includegraphics[width=\linewidth]{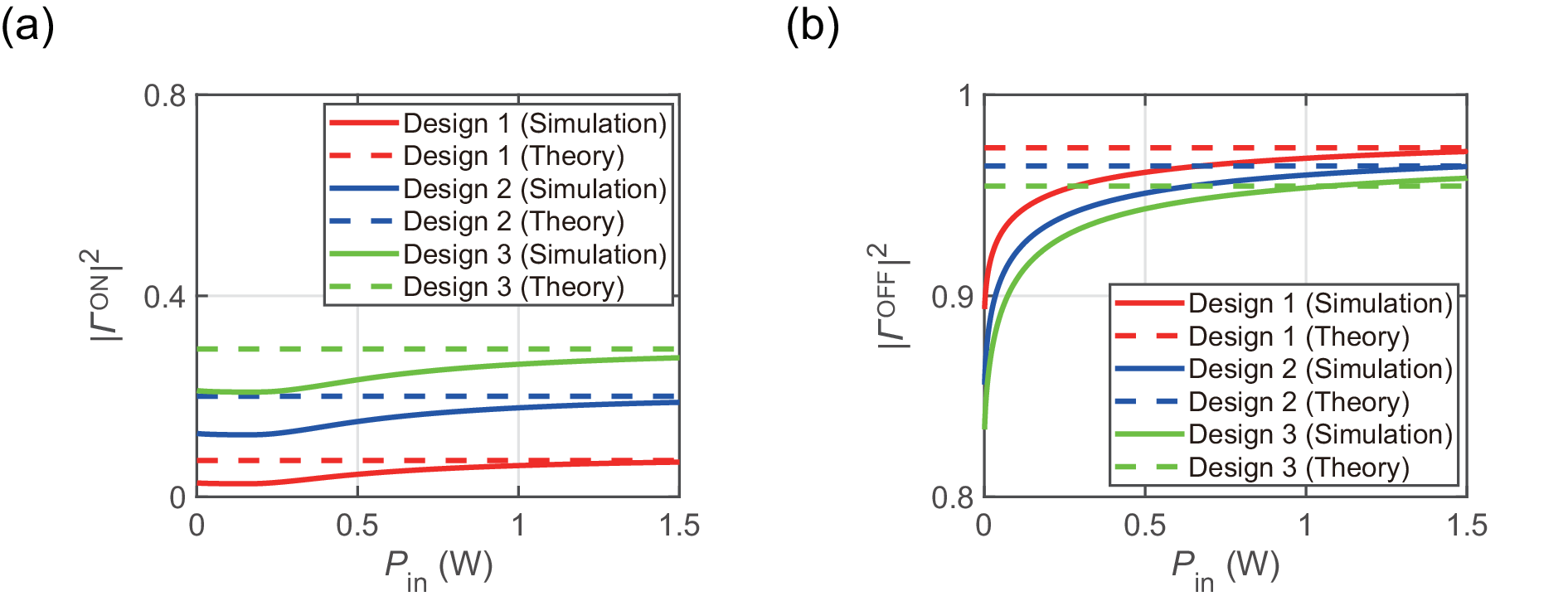}
\caption{Co-simulated and theoretical reflection magnitudes versus input power for Designs 1--3 in the (a) ON and (b) OFF states.}
\label{fig:9}
\end{figure}

The co-simulated reflection magnitudes are compared with the theoretical values in Fig. \ref{fig:9}. In the OFF state, all three designs maintain high reflectivity ($|\varGamma ^{\mathrm{OFF}}|^2 >$ 0.9) across the entire power range, which is consistent with the small losses in the high-impedance OFF-state MOSFET. In the ON state, the reflection magnitude increases with power for all designs and converges towards the theoretical prediction at high power. This convergence is expected because the analytical model employs the MOSFET parameters evaluated at the rated-current boundary (i.e., at the highest voltage within the safe operating range); thus, the agreement improves as the simulated operating point approaches this boundary. A systematic offset between theory and simulation persists even at high power; this is attributed to simplifications in the equivalent-circuit model, which neglects, for example, the patch-to-ground capacitance and the surface inductance of the metallic patch, both of which shift the effective resonance and alter the impedance matching condition.

\subsection*{Dynamic range evaluation}
The practical operating envelope of the proposed IRS is constrained from above by the maximum tolerable current in the MOSFET and from below by the minimum input power at which the binary reflection phase difference remains within an acceptable range. We evaluate both limits to define the dynamic range for each design.

\begin{figure}[ht!]
\centering
\includegraphics[width=0.5\linewidth]{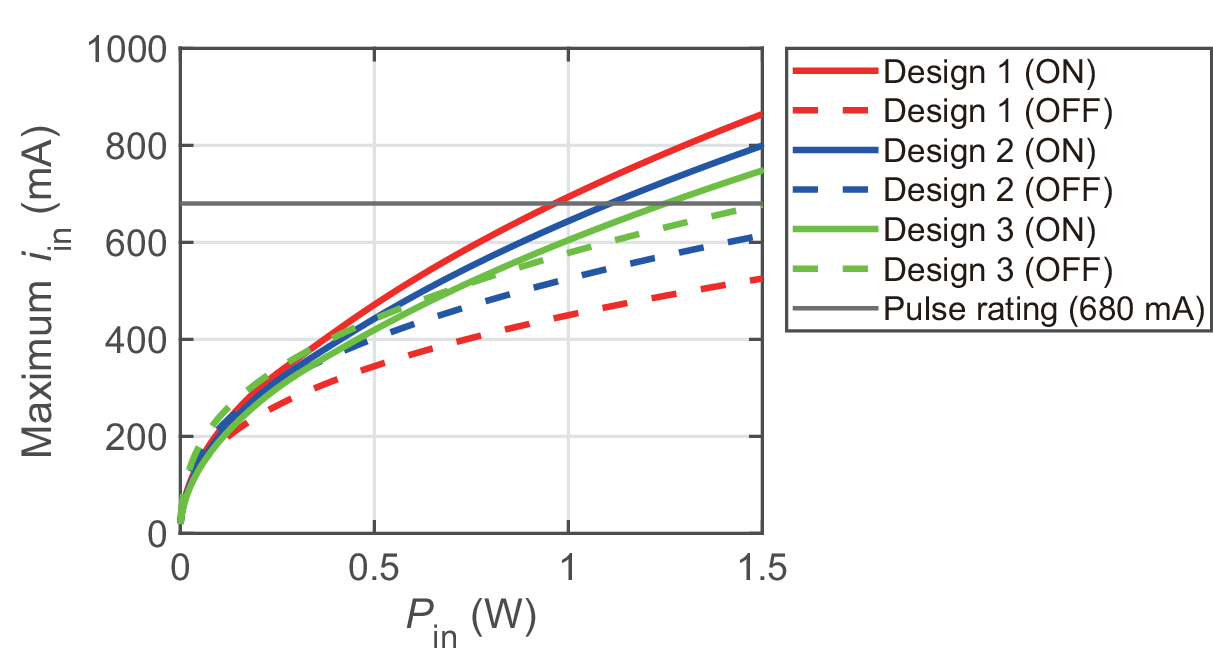}
\caption{Maximum current in the switching circuit as a function of input power for Designs 1--3 in the ON and OFF states, compared with the pulse-rated current of 680 mA. }
\label{fig:10}
\end{figure}

Fig. \ref{fig:10} shows the peak current flowing through the switching circuit as a function of $P_{\mathrm{in}}$ for Designs 1--3 in both the ON and OFF states. As expected from the impedance analysis, the ON-state current is consistently greater than the OFF-state current for all designs, and the upper power limit is therefore determined by the ON-state current reaching the 680-mA pulse rating. Among the three designs, Design 1 draws the highest current at any given power level because its larger $C_1$ and $C_2$ lower the overall circuit impedance seen by the induced gap voltage, thereby admitting more current into the switching circuit branch. The maximum permissible input powers, determined by interpolation of the current--power curves against the 680-mA threshold, are 0.965, 1.108, and 1.250 W for Designs 1, 2, and 3, respectively (Table \ref{tab:5}). Design 3 thus affords the highest upper bound on input power, a direct consequence of its smaller capacitances that limit the current drawn from the gap.

\begin{table}
    \centering
    \begin{tabular}{ccc}
    \hline
      Design   & Limiting state & Max. $P_{\mathrm{in}}$ (W)\\\hline\hline
       1  & ON & 0.965\\
       2  & ON & 1.108\\
       3  & ON & 1.250\\\hline
    \end{tabular}
    \caption{Dynamic range upper bound based on the rated current.}
    \label{tab:5}
\end{table}

\begin{figure}[ht!]
\centering
\includegraphics[width=\linewidth]{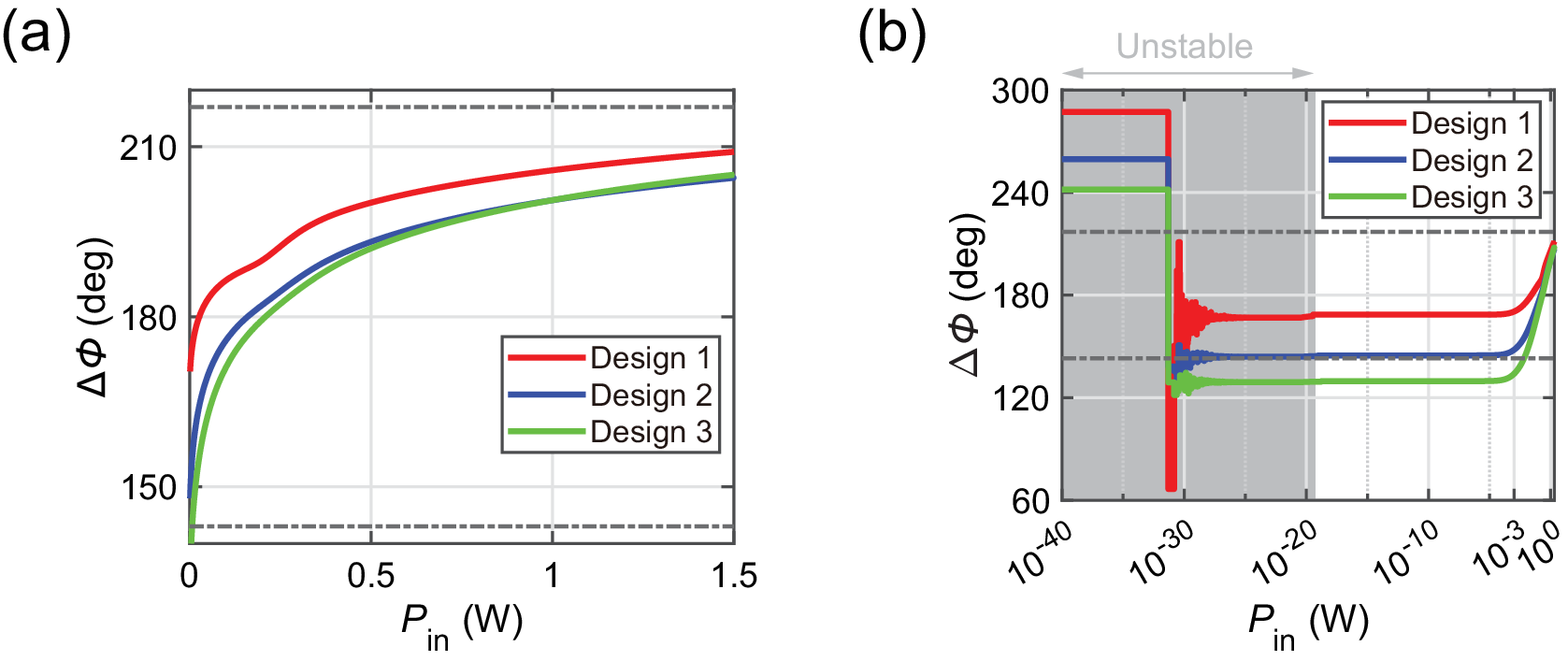}
\caption{Reflection phase difference $\Delta \varPhi$ between the ON and OFF states as a function of input power for Designs 1--3. (a, b) The results in (a) a linear scale and (b) a logarithmic scale. Dashed lines represent the acceptable range (180$^\circ \pm$ 37$^\circ$). The shaded region at extremely low power levels indicates numerical instability in the SPICE model.}
\label{fig:11}
\end{figure}

The lower bound of the dynamic range is governed by the reflection phase difference $\Delta \varPhi = \angle \varGamma ^{\mathrm{OFF}} - \angle \varGamma ^{\mathrm{ON}}$. For a 1-bit coded metasurface, a phase difference close to 180$^\circ$ is required for effective reflection control; following the criterion in the literature\cite{su2018uneven}, we adopt an acceptable range of $180^\circ \pm 37^\circ$, i.e., $143^\circ < \Delta \varPhi < 217^\circ$, with 143$^\circ$ as a conservative lower threshold. Fig. \ref{fig:11}a and Fig. \ref{fig:11}b show $\Delta \varPhi$ as a function of $P_{\mathrm{in}}$ for the three designs. Designs 1 and 2 satisfy the phase criterion over the entire simulated power range because their large $C_2$ stabilises both the ON- and OFF-state phases against low-power deviations. Design 3, in contrast, exhibits a phase difference dip below 143$^\circ$ at extremely low power ($P_{\mathrm{in}} <$ 0.006 W), imposing a lower bound on the usable power range. Note that by decreasing the input power to an extremely small value, $\Delta \varPhi$ goes beyond the range between $143^\circ$ and $217^\circ$ (Fig. \ref{fig:11}b). However, this deviation can be attributed to an instability issue in the numerical simulations as the drain--source current of the MOSFETs falls below the current range over which the SPICE model remains reliable, indicating that such low-power simulations are effectively improper. Moreover, Fig. \ref{fig:11}b shows that Designs 1 and 2 maintain $\Delta \varPhi$ between $143^\circ$ and $217^\circ$ even if $P_{in}$ is set to 10$^{-10}$ W, which is sufficiently low for WPT applications and even for wireless communications (namely, as a RIS).  Thus, effectively, Designs 1 and 2 satisfy the requirements for the lower and upper bounds of $\Delta \varPhi$. 

Table \ref{tab:6} combines the upper and lower bounds to give the full dynamic range for each design. Design 1 and Design 2 have no lower power limit (within the simulation range) but are constrained to 0.965 W and 1.108 W, respectively, at the upper end. Design 3 spans from 0.006 W to 1.250 W, offering the widest absolute dynamic range and the highest maximum operating power. The choice among these designs therefore depends on the application requirements: Design 1 is preferable when robustness against power fluctuations and manufacturing variability is required, whereas Design 3 is optimal when the highest possible power throughput is the primary objective.

\begin{table}
    \centering
    \begin{tabular}{ccc}
    \hline
      Design   & Min. $P_{\mathrm{in}}$ (W) & Max. $P_{\mathrm{in}}$ (W)\\\hline\hline
       1  & Not limited within the simulated range (lower than at least 10$^{-10}$) & 0.965\\
       2  & Not limited within the simulated range (lower than at least 10$^{-10}$) & 1.108\\
       3  & 0.006 & 1.250\\\hline
    \end{tabular}
    \caption{Combined dynamic range based on the rated current and reflection phase difference criteria.}
    \label{tab:6}
\end{table}

\subsection*{Supercell beam-steering demonstration}
To demonstrate that the proposed MOSFET-based IRS can perform wavefront engineering, we constructed finite supercell models and evaluated their far-field reflection patterns. The impedance sheets corresponding to the ON and OFF states of Design 3 at $P_{\mathrm{in}}$ = 1.250 W were extracted from the co-simulation data using the retrieval method\cite{smith2005electromagnetic, holloway2005reflection, fathnan2022method, kunitomo2024passive}, in which the metasurface is modelled as single-layer impedance sheets $Z_{mt}$s representing composite unit cells. For a grounded dielectric slab with no transmission, the composite surface impedance $Z_{es} = Z_0(1 + \varGamma)/(1 - \varGamma)$ includes propagation into the substrate, and the metallic pattern impedance is obtained by de-embedding as $Z_{mt} = jZ_{es}Z_s \tan{(\beta_s d)}/(jZ_s \tan{(\beta_s d)} - Z_{es})$, where $Z_s$ and $\beta_s$ are the substrate impedance and propagation constant, respectively.

\begin{figure}[ht!]
\centering
\includegraphics[width=\linewidth]{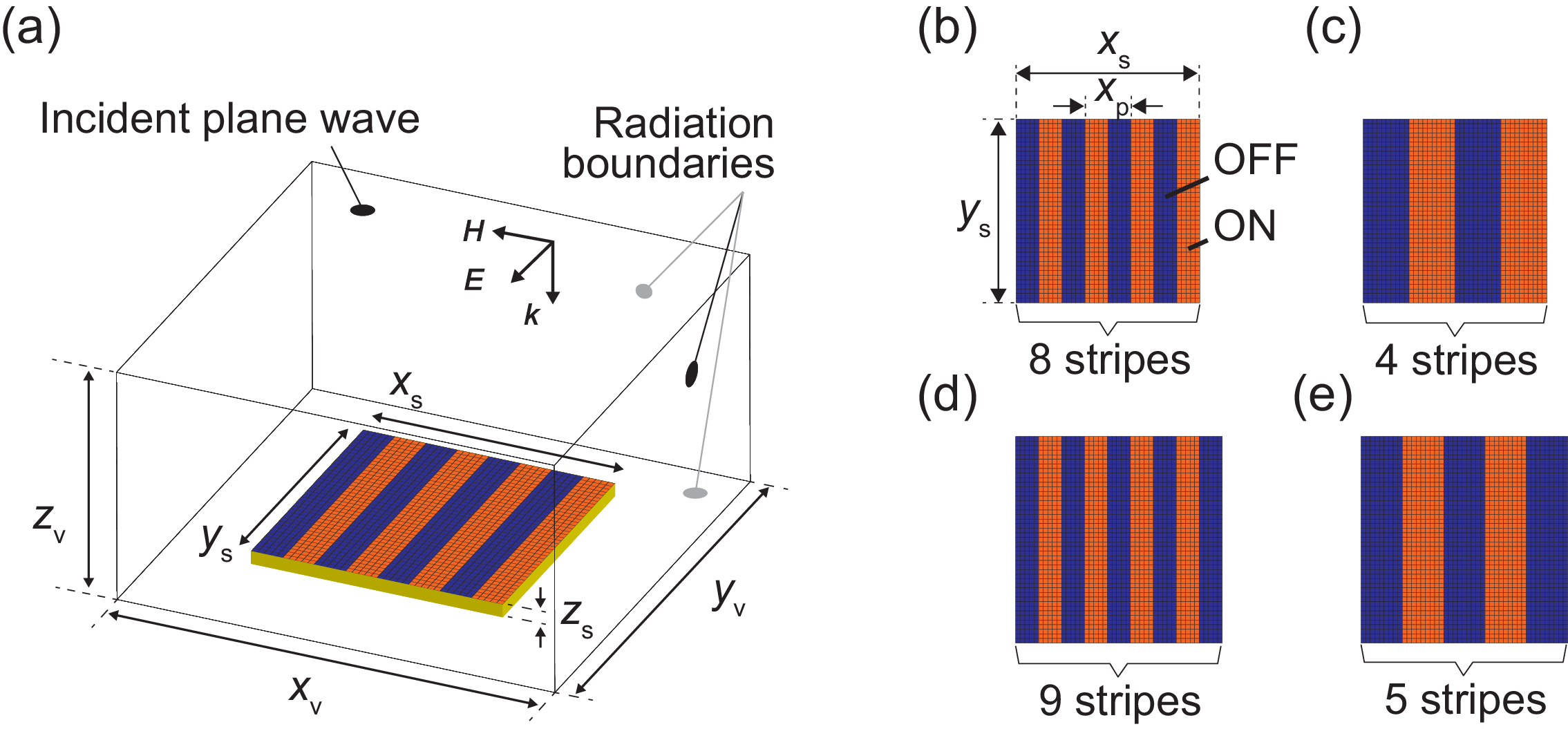}
\caption{Supercell models. (a) The entire simulation setup. Supercell models with (b) Pattern 1, (c) Pattern 2, (d) Pattern 3 and (e) Pattern 4 of unit cells. The orange and blue cells represent the ON- and OFF-state unit cells, respectively. }
\label{fig:12}
\end{figure}

The extracted ON and OFF impedance sheets were arranged in stripe patterns on planar substrates and illuminated from above by a normal incident wave at 2.4 GHz, with radiation boundary conditions enclosing the computational domain, as shown in Fig. \ref{fig:12}a. Two categories of supercell arrangements were investigated: asymmetric patterns (Patterns 1 and 2 drawn in Fig. \ref{fig:12}b and Fig. \ref{fig:12}c, respectively) on a 600 mm $\times$ 600 mm substrate and symmetric patterns (Patterns 3 and 4 drawn in Fig. \ref{fig:12}d and Fig. \ref{fig:12}e, respectively) on a 675 mm $\times$ 675 mm substrate. These four patterns were characterised by their supercell period $x_p$ and substrate width $x_s$. Pattern 1 employed alternating ON and OFF stripes of five unit-cell columns each, corresponding to a supercell period of $x_p =$ 150 mm. Four supercells therefore occupied the 600-mm aperture ($x_s =$ 600 mm), yielding eight stripes in total. Pattern 2 employed ten-column ON and OFF stripes, corresponding to $x_p =$ 300 mm, so that two supercells occupied the 600-mm aperture ($x_s =$ 600 mm) and yielded four stripes. Pattern 3 employed five-column stripes on a 675-mm aperture ($x_s =$ 675 mm), corresponding to $x_p =$ 150 mm; the aperture therefore accommodated 4.5 supercells and yielded a symmetric nine-stripe arrangement. Pattern 4 employed nine-column stripes, corresponding to $x_p =$ 270 mm; the aperture therefore accommodated 2.5 supercells and yielded a symmetric five-stripe arrangement ($x_s =$ 675 mm). The theoretical anomalous reflection angle under normal incidence for a stripe-coded binary metasurface is given by $\theta _{\mathrm{r}} = \sin^{-1}{(\lambda/x_p)}$\cite{cui2014coding, huang2022switchable}, where $\lambda$ is the free-space wavelength (2.4 GHz in this study). 

\begin{table}
    \centering
    \begin{tabular}{cccccccc}
    \hline
      Pattern   & $x_{v}$ (mm) & $y_{v}$ (mm)  & $z_{v}$ (mm) & $x_{s}$ (mm) & $y_{s}$ (mm) & $z_{s}$ (mm) & $x_{p}$ (mm)\\\hline\hline
      1   & 1100 & 1100  & 513.04 & 600 & 600 & 3.04 & 150\\
      2   & 1100 & 1100  & 513.04 & 600 & 600 & 3.04 & 300\\
      3   & 1175 & 1175  & 513.04 & 675 & 675 & 3.04 & 150\\
      4   & 1175 & 1175  & 513.04 & 675 & 675 & 3.04 & 270\\\hline\hline
    \end{tabular}
    \caption{Design parameters of the supercell models shown in Fig. \ref{fig:12}.}
    \label{tab:7}
\end{table}

\begin{figure}[ht!]
\centering
\includegraphics[width=\linewidth]{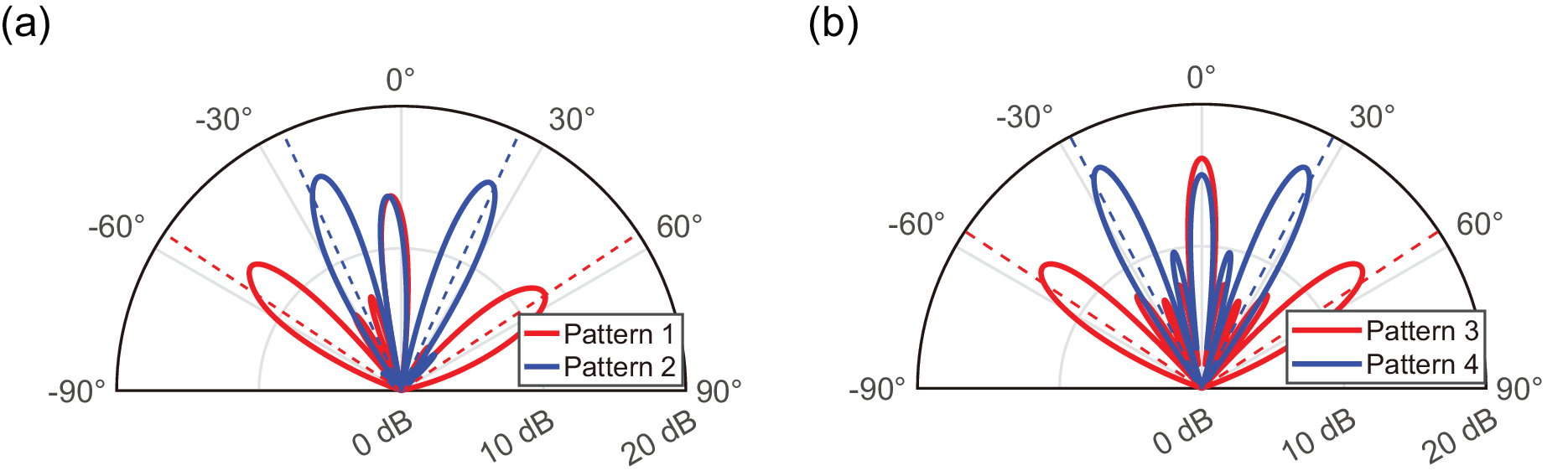}
\caption{Simulated far-field radiation patterns. (a, b) Results for (a) asymmetric and (b) symmetric supercells. The dashed lines represent the theoretical reflection angles.}
\label{fig:13}
\end{figure}

The simulated far-field radiation patterns for the asymmetric arrangements of Patterns 1 and 2 are shown in Fig. \ref{fig:13}a. Both patterns produce well-defined main lobes whose directions shift with the supercell period, qualitatively matching the theoretical prediction (dashed lines). However, the simulated beam directions are slightly shifted towards shallower angles on one side. This asymmetry and angular offset arise from the finite aperture: the theoretical expression of $\theta_r$ assumes an infinitely periodic array, whereas the 600-mm substrate supports only 4 (Pattern 1) or 2 (Pattern 2) supercell repetitions, leading to diffraction-related beam broadening and angular perturbation.

Fig. \ref{fig:13}b presents the corresponding results for the symmetric patterns. As expected from the symmetric cell layout, the far-field patterns exhibit bilateral symmetry: the main lobes appear at equal angles on either side of the broadside direction. Both patterns again deviate from the theoretical angles, which can be attributed to the finite number of noninteger supercell repetitions. Importantly, in all four patterns, the main-lobe direction changes predictably with the supercell period, confirming that the proposed MOSFET-based control circuit enables active electronic beam steering through reconfiguration of the binary cell pattern---without any physical modification to the metasurface---at an input power level of 1.25 W.

\section*{Discussion}
The results presented above demonstrate that the proposed MOSFET-based IRS operates stably at input power levels exceeding 1 W per unit cell at 2.4 GHz---a regime well beyond the reach of conventional PIN-diode or varactor-based designs, whose current ratings are typically limited to tens of milliamperes. The central design innovation lies in the combination of a back-to-back MOSFET switching topology, which provides bidirectional AC blocking, with a series--parallel capacitor stabilisation network that addresses two challenges simultaneously: it suppresses the voltage-dependent impedance nonlinearity inherent to MOSFETs at gigahertz frequencies, and it preserves the susceptance contrast needed for effective binary phase coding.

A principal insight from this work is the three-way trade-off among nonlinearity suppression, reflection efficiency, and dynamic range, mediated by the shunt inductance $L_{\mathrm{add}}$. Decreasing $L_{\mathrm{add}}$ permits a larger parallel capacitance $C_2$, which strengthens the stabilisation against device parameter variations and manufacturing tolerances. However, the enhanced robustness comes at the expense of a reduced ON-state reflection magnitude because the dominant $C_2$ homogenises the impedance contrast between the two states and shifts the ON-state impedance towards free-space matching, increasing absorption losses. Conversely, increasing $L_{\mathrm{add}}$ (as in Design 3) maximises both the reflection efficiency and the upper bound on the input power but yields a smaller $C_2$ and therefore greater sensitivity to nonlinear and manufacturing-induced impedance fluctuations. In practice, the optimal design point depends on the application context: Design 1 is suitable for scenarios where the ambient power density varies unpredictably or where large-scale arrays must tolerate significant interelement variability, whereas Design 3 is preferable when the transmitter power is tightly controlled and the primary objective is to maximise the reflected power.

The systematic discrepancy between the analytical transmission line model and the full-wave co-simulation highlights the limitations of the simplified equivalent circuit. The model captures the first-order behaviour---namely, the dependence of the reflection phase on the switching state, the ranking of designs by reflection magnitude, and the general trend of convergence at high power---but it underestimates the absolute phase values and overestimates the ON-state reflectivity. These errors originate from, for example, the neglect of the patch-to-ground capacitance (which lowers the effective resonant frequency) and the surface inductance of the metallic patch (which affects the impedance matching). Incorporating these parasitic elements into a higher-order equivalent circuit or calibrating the simplified model against a single full-wave simulation point would significantly improve the quantitative accuracy and could enable rapid design space exploration without repeated electromagnetic simulations.

Notably, the present design employs 1-bit (binary) phase coding, which inherently generates a specular reflection component in addition to the desired steered beam\cite{cui2014coding}. The energy distributed into the specular lobe reduces the gain of the anomalously reflected beam. To suppress this parasitic lobe, future implementations could adopt multibit phase quantisation---for instance, by incorporating two or more MOSFET branches with different circuit parameters to create four or more distinct impedance states. Alternatively, analogue phase tuning could be achieved by modulating the gate--source voltage of the MOSFET to continuously vary the channel conductance, although this approach would require a more sophisticated biasing network and careful linearisation of the phase--voltage characteristics.

In the future, the maximum operating power can be further increased through judicious device selection. Wide-bandgap semiconductors such as gallium nitride (GaN) high-electron-mobility transistors (HEMTs) can offer lower ON-resistance, higher breakdown voltage, and smaller output capacitance than the silicon MOSFET used here. These attributes would simultaneously widen the impedance contrast, increase the current rating, and reduce parasitic leakage, potentially pushing the dynamic range into the tens-of-watts regime. Our design framework, comprising the back-to-back topology, the $C_1$--$C_2$ stabilisation network, and the co-simulation validation workflow, is directly transferable to such advanced devices and provides a systematic methodology for high-power IRS development.

\section*{Conclusion}
We have proposed and validated a MOSFET-based IRS operating at 2.4 GHz for high-power wireless power transfer. Our findings yield the following design guidelines for high-power IRSs. First, a back-to-back MOSFET topology is essential for blocking bidirectional AC; selecting MOSFETs with low parasitic capacitance maximises the OFF-state impedance contrast at microwave frequencies. Second, the parallel capacitance across the switching circuit should be made as large as possible to suppress the impedance drift caused by device nonlinearity and manufacturing tolerance. Third, the resulting loss of susceptance contrast must be compensated by a correspondingly large series capacitance, with a shunt inductance providing resonance tuning while governing the trade-off between robustness and reflection efficiency. Fourth, the upper power limit is determined by the ON-state current; employing MOSFETs with higher rated currents and lower on-resistance directly extends the dynamic range. Applying these principles, we demonstrated stable beam steering up to 1.25 W, establishing a scalable framework for watt-level and beyond IRS-assisted WPT. The proposed framework provides a pathway towards the deployment of IRSs in practical high-power WPT scenarios.

\section*{Data availability}

The data that support the findings of this study are available within the article.


\providecommand{\noopsort}[1]{}\providecommand{\singleletter}[1]{#1}%

\section*{Acknowledgements}
This study was supported by the Japan Science and Technology Agency (JST) under Adopting Sustainable Partnerships for Innovative Research Ecosystem (ASPIRE) (Nos. JPMJAP2431 and JPMJAP2432) and Fusion Oriented Research for Disruptive Science and Technology (FOREST) (No. JPMJFR222T), KAKENHI grant from the Japan Society for the Promotion of Science (JSPS) (Nos. 23H00470 and 26H02139), and the Ministry of Internal Affairs and Communications (MIC) under the Fundamental Technologies for Sustainable Efficient Radio Wave Use R\&D Project (FORWARD) (No. JPMI250610002).

\section*{Contributions}

H.W. conceived and supervised the project. A.N. and G.I. designed the metasurfaces. G.I. conducted numerical simulations to validate the concept, with support from A.N. and E.O. All authors discussed the results and contributed to writing the manuscript.

\section*{Competing interests}

The authors declare that they have no competing interests.

\end{document}